\documentclass[a4paper,12pt]{article}
\usepackage{amsfonts}
\usepackage{amsmath}
\usepackage{amsthm}
\usepackage{amssymb}
\usepackage{url}
\usepackage{mathrsfs}
\usepackage{mathtools}
\usepackage{hhline}
\usepackage{ulem}
\usepackage{float}
\usepackage{bm}
\usepackage{textcomp}
\usepackage{enumitem}
\usepackage{physics}
\usepackage[longnamesfirst]{natbib}
\usepackage{xcolor}
\usepackage{booktabs}
\usepackage{hyperref}

\usepackage[margin=1.1in]{geometry}

\usepackage{setspace}

\setlist[itemize]{itemsep=0pt}
\setlist[enumerate]{itemsep=0pt}

\usepackage[mathlines]{lineno}
\usepackage{titlesec}

\titleformat{\section}
  {\normalsize\bfseries\centering} 
  {\thesection.} 
  {0.5em} 
  {} 

\titleformat{\subsection}
  {\normalfont\centering} 
  {\thesubsection.}    
  {0.5em}              
  {\itshape}    

\titleformat{\subsubsection}
  {\normalfont\centering} 
  {\thesubsubsection}    
  {0.5em}              
  {}    

\titleformat{name=\section, numberless}
  {\normalsize\centering\scshape}
  {}
  {0em}
  {}

\theoremstyle{definition}
\newtheorem{definition}{Definition}

\newtheorem{example}{Example}

\theoremstyle{plain}
\newtheorem{lemma}{Lemma}

\newtheorem{theorem}{Theorem}
\newtheorem{proposition}{Proposition}
\newtheorem{corollary}{Corollary}

\mathtoolsset{showonlyrefs=true}

\title{Obvious Group Manipulations\\
and Top Trading Cycles\thanks{Doi acknowledges financial support from JST SPRING Grant Number JPMJSP2123.}}

\author{
Ryoga Doi\thanks{Graduate School of Economics, Keio University,
Tokyo 108-8345, Japan. Email: \mbox{\href{mailto:doiryouga@keio.jp}{doiryouga@keio.jp}}} ~~
Kazumasa Ikeda\thanks{Graduate School of Economics, Keio University,
Tokyo 108-8345, Japan. Email: \mbox{\href{mailto:ikeda-k25@keio.jp}{ikeda-k25@keio.jp}}} ~~
Kensei Nakamura\thanks{Graduate School of Economics, Hitotsubashi University, Tokyo 186-0004, Japan. Email: \mbox{\href{mailto:kensei.nakamura.econ@gmail.com}{kensei.nakamura.econ@gmail.com}}}}

\begin{document}

\maketitle

\begin{abstract}
\noindent
    We introduce a group version of non-obvious manipulability, which is stronger than the original definition by \citet{troyan2020obvious} and weaker than group strategy-proofness, and study its implications in Shapley--Scarf housing markets.
    We provide several characterizations of the Top Trading Cycles mechanism using the new condition. 
    \vspace{1mm}
\end{abstract}

\section{Introduction}
\label{sec:intro}

This paper studies the Shapley--Scarf housing market \citep{shapley1974cores} from the perspective of invulnerability to strategic manipulation.
Famously, as shown by \citet{roth1982incentive}, the Top Trading Cycles mechanism (henceforth, TTC) satisfies strategy-proofness, which requires that no agent can obtain a more preferred object by misreporting their preferences.
Moreover, under Pareto efficiency and individual rationality, TTC is the unique strategy-proof mechanism \citep{ma1994strategy}.
Although strategy-proofness is a cornerstone for evaluating mechanisms in terms of the manipulability of outcomes, it is not the only relevant notion.
The best-known stronger notion is group strategy-proofness, which accounts for strategic manipulation not only by individual agents but also by groups of agents.
It is known that TTC satisfies this stronger property as well \citep{bird1984group}.\footnote{For recent surveys of TTC, see \citet{afacan2024housing} and \citet{morrill2024top}.} 

After decades of research on strategy-proofness, two important new concepts have recently been introduced.
The first is the stronger condition called obvious strategy-proofness \citep{li2017obviously}.
Loosely speaking, it requires agents, at every point of some extensive-form mechanism, to have an incentive to truthfully report their preferences even if they pessimistically predict the outcome following a truthful report and optimistically predict the outcome following a misreport.
Although TTC admits no obviously strategy-proof implementation in the standard setting, it does admit one on the domain of single-dipped preferences \citep{tamura2023object}. A different mechanism, known as the crawler, admits such an implementation on the domain of single-peaked preferences \citep{bade2019matching}.

The second is a weakening of strategy-proofness that is motivated by the need to evaluate how vulnerable non-strategy-proof mechanisms are to manipulation \citep{troyan2020obvious}.
A mechanism is said to be \textit{not obviously manipulable (NOM)} if no agent has a misreport that improves, relative to truthful reporting, either the best scenario (i.e., the outcome obtained when all other agents act in the way most favorable to that agent) or the worst scenario (i.e., the outcome obtained when all other agents act in the way least favorable to that agent). 
\citeauthor{troyan2020obvious} showed that this property is equivalent to the absence of manipulations that cognitively limited agents can recognize as profitable, and examined it in several models, such as school choice and auctions.
This concept has been studied in many papers, such as \citet{ortega2022obvious}, \citet{troyan2024obvious}, \citet{arribillaga2024obvious,arribillaga_bonifacio2025_ET}, \citet{arribillaga_risma2025}, and \citet{de2026breaking}.

To obtain a better understanding of strategic manipulability in housing markets, this paper studies NOM and its group variant on several domains of preference profiles, such as the full domain, the single-peaked domain, and the single-dipped domain.
We first show that, under a very weak condition on the preference domain, every pair efficient and individually rational mechanism is NOM.
Note that pair efficiency is a condition much weaker than the standard Pareto efficiency condition \citep{ekici2024pair}. 
It requires that no two agents can both benefit by swapping their assigned objects.
Thus, roughly speaking, this result suggests that NOM has little power to distinguish among mechanisms, even when attention is restricted to well-known preference subdomains.

Motivated by this observation, we introduce a stronger notion that accounts for group misreporting. 
A mechanism is said to be \textit{not obviously group manipulable (NOGM)} if no group of agents has a weak incentive to misreport, in the same spirit as NOM.
We define NOGM so that it is a natural extension of NOM in light of the characterization by \citet{troyan2020obvious}; that is, just as NOM is equivalent to the absence of manipulations that cognitively limited agents can recognize as profitable, NOGM is equivalent to the absence of such manipulations by groups of cognitively limited agents.
NOGM is weaker than group strategy-proofness but stronger than NOM.\footnote{
\label{fn:otherNOGM}
Note that \cite{Bonifacio2026} also introduced a different group version of NOM. 
% Their condition is stronger than group strategy-proofness.
The key difference lies in how the best and worst outcomes are defined. They define these outcomes separately for each member of the coalition, according to her own preference. In contrast, we define them jointly for the coalition: the best (resp. worst) outcomes are the achievable allocations that are not Pareto dominated by (resp. do not Pareto dominate) any other achievable allocation for the members of the coalition.
} 
We characterize TTC by NOGM, individual rationality, and pair efficiency on the full domain and on the single-dipped domain.
Furthermore, if we focus on the full domain, we can replace pair efficiency with the ontoness of the mechanism in our characterization.

To clarify the contribution of our characterizations, we briefly discuss the related literature. 
\citet{ekici2024pair} characterized TTC on the full domain using pair efficiency, individual rationality, and strategy-proofness; that is, Ekici replaced Pareto efficiency with its pair version in \citeauthor{ma1994strategy}'s (\citeyear{ma1994strategy}) characterization. 
Relative to \citeauthor{ekici2024pair}'s result, our full-domain characterization replaces strategy-proofness with NOGM (note that the two conditions are logically independent).
Moreover, \citet{takamiya2001coalition} characterized TTC by ontoness, individual rationality, and \citeauthor{maskin1999nash}'s (\citeyear{maskin1999nash}) monotonicity, which is equivalent to group strategy-proofness in housing markets.
Our result shows that, on the full domain, group strategy-proofness in this characterization can be weakened to NOGM.

On the single-dipped domain, \citet{tamura2023object} showed that \citeauthor{ma1994strategy}'s characterization remains valid, and \citet{hu2024characterization} showed the same for \citeauthor{ekici2024pair}'s characterization; that is, both characterize TTC using strategy-proofness.
In contrast, our single-dipped characterization uses NOGM instead of strategy-proofness.

Since \citet{ekici2024pair}, pair efficiency has  been studied intensively (e.g., \citealp{ekici2024characterizing,hu2024characterization,goel2025ttc,mandal2025efficient,klaus2025characterization}). 
Many of these papers study its implications in combination with strategy-proofness or stronger incentive conditions.
This paper contributes to this literature by examining pair efficiency under weakenings of (group) strategy-proofness.

Finally, we note that \citet{demoor2026non} is an independent work that also studies obvious manipulations by groups in housing markets. 
Their analysis is motivated by a result stating that any Pareto efficient and individually rational mechanism on the full domain is NOM. 
Although we show a similar implication using pair efficiency instead of Pareto efficiency, this difference is not essential (see the discussion in Section~\ref{subsec:ind}). 
The important difference is the domain assumption: we show a similar property under a very mild assumption on the preference domain. 
They also study a group version of NOM, but their definition is quite different from ours.
They focus on misreporting by a single agent who knows the strategy chosen by the other members in an exogenously fixed group, and study properties of mechanisms different from the standard TTC.

This paper is organized as follows.
Section~\ref{sec:model} introduces the formal model and basic properties of mechanisms.
Section~\ref{sec:manip} examines the implications of NOM and NOGM. 
Section~\ref{sec:con} concludes.
In Appendix~\hyperref[app:A]{A}, we provide a characterization of  NOGM within a general framework, in the spirit of \citet{troyan2020obvious}. 
All proofs of the results in the main text are in Appendix~\hyperref[app:B]{B}.
Appendix~\hyperref[app:C]{C} shows that strategy-proofness and NOGM are logically independent and group strategy-proofness implies NOGM.

\section{Model}
\label{sec:model}

We adopt the standard model of housing markets \citep{shapley1974cores}. 
Let $N = \{1,2,\ldots, n\}$ be the set of agents, with $n\ge 3$. %
% \footnote{
% \textcolor{red}{How about the case with $n=2$?} 
% } 
Let $O = \{o_1 , \ldots, o_n \}$ be the set of indivisible objects, where each $o_i$ is agent $i$'s endowment. 
An \textit{allocation} is a one-to-one mapping $a : N \to O$. 
For notational simplicity, we write $a_i$ for agent $i$'s assignment at $a$, that is, $a_i = a(i)$.
The set of allocations is denoted by $\mathcal{A}$.

Each agent has a strict preference relation $\succ_i$ over objects. 
For $o, o' \in O$, $o \succ_i o'$ means that  $i$ strictly prefers $o$ to $o'$. 
We write $o \succeq_i o'$ if and only if $o \succ_i o'$ or $o = o'$.
Let $\mathcal{P}$ be the set of strict preferences.
We denote by $\succ = (\succ_i)_{i\in N} \in \mathcal{P}^N$ the preference profile of the agents in $N$.

% \textcolor{red}{TBA: the definitions of $\succ_{-i}$, $\succ_{-S}$, $a_{-i}$, and so on}
For $\succ \in \mathcal{P}^N$ and $S\subseteq N$, we write $\succ_S=(\succ_i)_{i\in S}\in \mathcal{P}^S$ for the subprofile of the agents in $S$, and $\succ_{-S}=\succ_{N\setminus S}\in \mathcal{P}^{N\setminus S}$ for that of the others.
When $S=\{i\}$, we write $\succ_{-i}$ instead of $\succ_{-\{i\}}$. 
For $\succ'_S\in \mathcal{P}^S$ and $\succ_{-S}\in \mathcal{P}^{N\setminus S}$, $(\succ'_S,\succ_{-S})$ denotes the profile in $\mathcal{P}^N$ whose restrictions to $S$ and to $N\setminus S$ are $\succ'_S$ and $\succ_{-S}$, respectively. 
We identify $(\succ'_N,\succ_{-N})$ with $\succ'_N$. 
Finally, once subprofiles $\succ'_S$ and $\succ'_{-S}$ have been specified, we write $\succ'$ for $(\succ'_S,\succ'_{-S})$.

Let $\mathcal{D}$ be a nonempty subset of $\mathcal{P}$. 
A \textit{mechanism} $\varphi$ is a function from $\mathcal{D}^N$ to $\mathcal{A}$. It assigns an allocation to each preference profile.
For each $i\in N$, let $\varphi_i (\succ)$ denote agent $i$'s object under $\varphi (\succ)$, that is, $\varphi_i (\succ) = a_i$, where $a = \varphi (\succ)$. 
We sometimes omit specifying the domain of a mechanism. 

% In this paper, we mainly consider three domains of mechanisms. 
Here, we introduce three representative domains. 
The first is the set of profiles of strict preference relations, that is, $\mathcal{P}^N$. 
The second and third are single-peaked and single-dipped domains, respectively. 
Suppose that there is a strict order $\triangleright$ over $O$.
For $o, o' \in O$, we write $o \trianglerighteq o'$ if $o \triangleright o'$ or $o = o'$.
Without loss of generality, we assume that $o_1\triangleright o_2 \triangleright\cdots \triangleright o_n$. 
We say that a preference $\succ_i \in \mathcal{P}$ is \textit{single-peaked} if 
there is $o_k\in O$ such that, for any $o_j,o_\ell\in O$,
\begin{equation}
    [o_k \trianglerighteq o_j \triangleright o_\ell \quad \text{or} \quad o_\ell \triangleright o_j \trianglerighteq o_k] \implies o_j \succ_i o_\ell.
\end{equation}
That is, the preference ranking rises monotonically toward the peak $o_k$ along the order $\triangleright$, and falls monotonically as one moves away from it on either side.
The set of single-peaked preferences is denoted by $\hat{\mathcal{P}}$. 
We say that a preference $\succ_i \in \mathcal{P}$ is \textit{single-dipped} if there exists $o_k\in O$ such that, for any $o_j,o_\ell\in O$,
\begin{equation}
    [o_k \trianglerighteq o_j \triangleright o_\ell \quad \text{or} \quad o_\ell \triangleright o_j \trianglerighteq o_k] \implies o_\ell \succ_i o_j.
\end{equation}
That is, the preference ranking falls monotonically as one approaches the dip $o_k$ along the order $\triangleright$, and rises monotonically as one moves away from it.
The set of single-dipped preferences is denoted by $\check{\mathcal{P}}$.

We say that $\mathcal{D}$ satisfies \textit{top-attainability} if, for every object $o\in O$, there exists a preference relation in $\mathcal{D}$ that ranks $o$ first.
Similarly, $\mathcal{D}$ satisfies \textit{bottom-attainability} if, for every object $o\in O$, there exists a preference relation in $\mathcal{D}$ that ranks $o$ last.
Note that $\hat{\mathcal{P}}$ satisfies top-attainability and that $\check{\mathcal{P}}$ satisfies bottom-attainability.

We introduce several properties of mechanisms. 
A mechanism $\varphi$ is \textbf{individually rational} if, for all $\succ\in \mathcal{D}^N$ and all $i\in N$, $\varphi_i (\succ) \succeq_i o_i$. 
An allocation $a\in \mathcal A$ is \textit{Pareto efficient} at $\succ\in \mathcal{P}^N$ if there is no allocation $a'$ such that $a'_i \succeq_i a_i$ for all $i\in N$ and $a'_j \succ_j a_j$ for some $j \in N$. 
A mechanism $\varphi$ is \textbf{Pareto efficient} if $\varphi (\succ)$ is Pareto efficient at $\succ$ for all $\succ\in \mathcal{D}^N$. 
We also study a weaker efficiency condition proposed by \citet{ekici2024pair}. 
An allocation $a\in \mathcal A$ is \textit{pair efficient} at $\succ\in \mathcal{P}^N$ if there is no pair $(i, j)\in N\times N$ such that $a_j \succ_i a_i$ and $a_i \succ_j a_j$.
That is, pair efficiency means that no pair of agents can both benefit by exchanging their assigned objects.
A mechanism $\varphi$ is \textbf{pair efficient} if $\varphi (\succ)$ is pair efficient at $\succ$ for all $\succ\in \mathcal{D}^N$. 
Moreover, a mechanism $\varphi$ is \textbf{onto} if, for all $a\in \mathcal{A}$, there exists $\succ\in \mathcal{D}^N$ such that $\varphi (\succ) = a$. 
If $\mathcal{D}$ satisfies top-attainability, then ontoness is implied by Pareto efficiency. 

As discussed in the Introduction, our main focus is on invulnerability to strategic manipulation.
A mechanism $\varphi$ is \textbf{strategy-proof} if, for all $\succ \in \mathcal{D}^N$, all $i\in N$, and all $\succ_i' \in \mathcal{D}$,  $\varphi_i (\succ) \succeq_i \varphi_i (\succ_i', \succ_{-i})$. 
Furthermore, a mechanism $\varphi$ is \textbf{group strategy-proof} if, for all $\succ\in \mathcal{D}^N$ and all $S\subseteq N$, there is no $\succ'_S \in \mathcal{D}^S$ such that $\varphi_i (\succ'_S, \succ_{-S}) \succeq_i \varphi_i ( \succ ) $ for all $i\in S$ and  $\varphi_j (\succ'_S, \succ_{-S}) \succ_j \varphi_j ( \succ ) $ for some $j\in S$. 
By definition, group strategy-proofness implies strategy-proofness.

The most popular mechanism is the TTC mechanism, which \citet{shapley1974cores} attribute to David Gale.
Formally, a mechanism is \textit{TTC} if it assigns objects to agents by the following algorithm. 
 
\begin{quote}
\textbf{TTC algorithm.} 

\noindent
\textbf{Step 1.} 
Each agent points to the agent who owns her most preferred object in $O$. 
There exists at least one cycle.
For each cycle, assign each agent in the cycle the object owned by the agent she points to, that is, \textit{execute} the cycle, and remove all such agents and objects from the market.

\medskip

\noindent
\textbf{Step $k$ ($k \ge 2$).} 
Let $N^{(k-1)}$ and $O^{(k-1)}$ be the sets of agents and objects remaining after Step $k-1$. 
Each agent in $N^{(k-1)}$ points to the agent who owns her most preferred object in $O^{(k-1)}$. 
There exists at least one cycle. Execute all the cycles, and remove all such agents and objects from the market.

\medskip

The procedure terminates in a finite number of steps when no agent remains.
\end{quote}

It is known that TTC has many desirable properties. 
For $\succ\in\mathcal{P}^N$, we say that a coalition $S\subseteq N$ \textit{blocks} an allocation $a$ if there exists an allocation $a'$ such that $\{a'_k \mid k\in S\} = \{o_k \mid k\in S\}$, $a'_i  \succeq_i a_i$ for all $i\in S$, and $a'_j \succ_j a_j$ for some $j\in S$. 
The \textit{(strict) core} is the set of allocations that are not blocked by any coalition.
\citet{roth1977weak} showed that, for all $\succ\in \mathcal{P}^N$, the core is a singleton consisting of the TTC allocation.
For axiomatic characterizations of TTC, see the papers mentioned in Section~\ref{sec:intro}. 

% As shown by \citet{ma1994strategy}, TTC is the unique mechanism on $\mathcal{P}^N$ that is individually rational, Pareto efficient, and strategy-proof. 
% \citet{ekici2024pair} proved that in this characterization, Pareto efficiency can be replaced with pair efficiency. 
% \citet{takamiya2001coalition} showed that TTC is characterized by individual rationality, ontoness, and group strategy-proofness on the domain $\mathcal{P}^N$. 
% See Section~\ref{subsec:lite} for other characterizations of TTC in the domains $\mathcal{P}^N$ and $\hat{\mathcal{P}}^N$.

\section{Obvious Manipulability}
\label{sec:manip}

\subsection{Obvious Manipulations by a Single Agent}
\label{subsec:ind}

Following \citet{troyan2020obvious}, we consider a weak version of strategy-proofness.
Given a mechanism $\varphi$, $i\in N$, and $\succ_i,\succ'_i\in\mathcal{D}$, we say that an allocation $b\in \{\varphi(\succ'_i, \succ_{-i}) \mid \succ_{-i}\in\mathcal{D}^{N\setminus\{i\}}\}$ is a \textbf{best allocation} at $(\varphi,\succ_i,\succ'_i)$ if there exists no $\succ_{-i}\in\mathcal{D}^{N\setminus\{i\}}$ such that $ \varphi_i(\succ'_i, \succ_{-i}) \succ_i b_i$. 
Similarly,  $w\in \{\varphi(\succ'_i, \succ_{-i}) \mid \succ_{-i}\in\mathcal{D}^{N\setminus\{i\}}\}$ is a \textbf{worst allocation} at $(\varphi,\succ_i,\succ'_i)$ if there exists no $\succ_{-i}\in\mathcal{D}^{N\setminus\{i\}}$ such that $w_i \succ_i \varphi_i(\succ'_i, \succ_{-i})$. 
Let $B^\varphi_{\succ_i}(\succ'_i)$ (resp. $W^\varphi_{\succ_i}(\succ'_i)$) denote the set of all best (resp.~worst) allocations at $(\varphi,\succ_i,\succ'_i)$. 
Note that for all $b,b'\in B^\varphi_{\succ_i}(\succ'_i)$ (resp.~$w,w'\in W^\varphi_{\succ_i}(\succ'_i)$), $b_i=b'_i$ (resp.~$w_i=w'_i$). 
A mechanism $\varphi$ is \textbf{not obviously manipulable (NOM)} if, for all $i\in N$, all $\succ_i\in\mathcal{D}$, all $w\in W_{\succ_i}^\varphi(\succ_i)$, and all $b\in B_{\succ_i}^\varphi(\succ_i)$, (i) there exists no $\succ'_i \in \mathcal{D}$ such that for all $w'\in W_{\succ_i}^\varphi(\succ'_i)$, $w'_i \succ_i w_i$, and (ii) there exists no $\succ'_i \in \mathcal{D}$ such that for all $b'\in B_{\succ_i}^\varphi(\succ'_i)$, $b'_i \succ_i b_i$.
In words, this condition requires that agents cannot improve the best or worst outcome by misreporting. 
\citeauthor{troyan2020obvious} showed that this condition can be characterized as strategy-proofness for agents who may not fully understand the mechanism. 
They examined the implications of this property in several models, such as school choice, auctions, and bilateral trade.

The next result shows that this condition is too weak in the Shapley--Scarf housing market. 

\begin{proposition}
\label{prop:non-restriction}
    Suppose that $\mathcal{D}$ satisfies top-attainability or bottom-attainability.
    Then every individually rational and pair efficient mechanism on $\mathcal{D}^N$ is NOM. 
\end{proposition}

In words, this result shows that under a mild domain assumption, NOM cannot distinguish among reasonable mechanisms in terms of manipulability. 

A similar result has been established independently by \citet{demoor2026non}.
They show that on the full domain, if a mechanism satisfies individual rationality and Pareto efficiency, then it is NOM. 
Note that the difference in the efficiency properties is not essential: 
it is easy to check that Pareto efficiency can be replaced with pair efficiency in their theorem. 
Rather, the crucial difference is in the domain assumption. 
Our proposition shows that even if the preference domain is restricted, NOM remains too weak in housing markets. 

\subsection{Obvious Manipulations by Groups}

Motivated by the weakness of NOM identified in Proposition~\ref{prop:non-restriction}, we introduce a stronger version of NOM in the spirit of group strategy-proofness. 

Given a mechanism $\varphi$, $S\subseteq N$, and $\succ_S,\succ'_S\in\mathcal{D}^S$, we say that an allocation $b\in \{\varphi(\succ'_S, \succ_{-S}) \mid \succ_{-S}\in\mathcal{D}^{N\setminus S}\}$ is a \textbf{best allocation} at $(\varphi,\succ_S,\succ'_S)$ if there exists no $\succ_{-S}\in\mathcal{D}^{N\setminus S}$ such that $ \varphi_i (\succ'_S, \succ_{-S}) \succeq_i b_i$ for all $i\in S$ and $\varphi_j (\succ'_S, \succ_{-S}) \succ_j b_j $ for some $j\in S$. 
Similarly,  $w\in \{\varphi(\succ'_S, \succ_{-S}) \mid \succ_{-S}\in\mathcal{D}^{N\setminus S}\}$ is a \textbf{worst allocation} at $(\varphi,\succ_S,\succ'_S)$ if there exists no $\succ_{-S}\in\mathcal{D}^{N\setminus S}$ such that  $w_i \succeq_i \varphi_i (\succ'_S, \succ_{-S})$ for all $i\in S$ and $w_j \succ_j \varphi_j (\succ'_S, \succ_{-S})$ for some $j\in S$.
Let $B^\varphi_{\succ_S}(\succ'_S)$ (resp. $W^\varphi_{\succ_S}(\succ'_S)$) denote the set of all best (resp.~worst) allocations at $(\varphi,\succ_S,\succ'_S)$. 

That is, $B^\varphi_{\succ_S} (\succ'_S)$ collects allocations that are achievable under the report $\succ'_S$ and Pareto undominated for $S$ among such allocations with respect to the true preference subprofile $\succ_S$. 
Similarly, $W^\varphi_{\succ_S} (\succ'_S)$ collects allocations that are achievable under the report $\succ'_S$ and do not Pareto dominate any other such allocation for $S$ with respect to the true preference subprofile $\succ_S$.

% let $B^\varphi_{\succ_S} (\succ'_S)$ be the set of allocations $b\in \{\varphi (\succ'_S, \succ_{-S} ) \mid \succ_{-S} \in \mathcal{D}^{N\setminus S} \}$ such that 
% there is no $\succ'_{-S}\in \mathcal{D}^{N\setminus S}$ with $ \varphi_i (\succ'_S, \succ'_{-S}) \succeq_i b_i$ for all $i\in S$ and $\varphi_j (\succ'_S, \succ'_{-S}) \succ_j b_j $ for some $j\in S$.

% Similarly, let $W^\varphi_{\succ_S} (\succ'_S)$ be the set of allocations $w\in \{\varphi (\succ'_S, \succ_{-S} ) \mid \succ_{-S} \in \mathcal{D}^{N\setminus S} \}$ such that 
% there is no $\succ'_{-S}\in \mathcal{D}^{N\setminus S}$ with $w_i \succeq_i \varphi_i (\succ'_S, \succ'_{-S})$ for all $i\in S$ and $w_j \succ_j \varphi_j (\succ'_S, \succ'_{-S})$ for some $j\in S$.

Using these two sets, we define a group version of NOM.

\begin{definition}
    Given a mechanism $\varphi$, $S\subseteq N$, and  $\succ_S\in \mathcal{D}^S$, we say that  $\succ'_S\in \mathcal{D}^S$ is an \textbf{obvious group manipulation} for $(\succ_S,\varphi)$ if either 
    \begin{itemize}
        \item[(i)] there exists $b'\in B^\varphi_{\succ_S}(\succ'_S)$ such that, for all $b\in B^\varphi_{\succ_S}(\succ_S)$,     $b'_i \succeq_i b_i$ for all $i\in S$ and $b'_j \succ_j b_j$ for some $j\in S$; or
        \item[(ii)] there exists $w\in W^\varphi_{\succ_S}(\succ_S)$ such that, for all $w'\in W^\varphi_{\succ_S}(\succ'_S)$,  
    $w'_i \succeq_i w_i $ for all $i\in S$ and  $w'_j  \succ_j w_j$ for some $j\in S$.
    \end{itemize}
\end{definition}

We say that a mechanism $\varphi$ is \textbf{not obviously group manipulable (NOGM)} if, for all $S\subseteq N$ and all $\succ_S\in\mathcal{D}^S$, there is no $\succ'_S\in\mathcal{D}^S$ that is an obvious group manipulation for $(\succ_S,\varphi)$.
This is a natural extension of NOM.
Indeed, in Appendix~\hyperref[app:A]{A}, we show that NOGM can be characterized as group strategy-proofness under mechanism uncertainty.
This is the counterpart of the characterization of NOM in \citet{troyan2020obvious}.%
\footnote{See also Footnote~\ref{fn:otherNOGM}.}
Note that NOGM is stronger than NOM and weaker than group strategy-proofness, but is logically independent of strategy-proofness (see Appendix~\hyperref[app:C]{C} for details).

The following theorem shows that NOGM characterizes TTC together with basic axioms on the full domain.

\begin{theorem}
\label{thm:TTC_full}
    Let $\varphi : \mathcal{P}^N \to \mathcal{A}$ be a mechanism.
    The following three statements are equivalent:
    \begin{itemize}
        \item[(i)] $\varphi$ is individually rational, pair efficient, and NOGM.
        \item[(ii)] $\varphi$ is individually rational, onto, and NOGM.
        \item[(iii)] $\varphi$ is TTC.
    \end{itemize}
\end{theorem}

This result suggests that, unlike NOM, its group version is strong enough to classify mechanisms in terms of strategic manipulability, and that TTC is again singled out when combined with natural properties. 

Related results are known in the literature. 
As mentioned in Section~\ref{sec:intro}, \citet{ekici2024pair} characterized TTC using individual rationality, pair efficiency, and strategy-proofness. 
Compared with \citeauthor{ekici2024pair}'s theorem, our theorem shows that strategy-proofness can be replaced with NOGM, which is logically independent of strategy-proofness (see Appendix~\hyperref[app:C]{C} for details). 
Furthermore, \citet{takamiya2001coalition} provided a characterization of TTC using individual rationality, ontoness, and group strategy-proofness.\footnote{
Strictly speaking, this result is a conjunction of two results by \citet{takamiya2001coalition}: 
TTC is the unique mechanism that is individually rational, onto, and monotone in the sense of \citet{maskin1999nash}, and monotonicity is equivalent to group strategy-proofness.
}
Our theorem shows that in \citeauthor{takamiya2001coalition}'s result, group strategy-proofness can be replaced with a weaker condition, NOGM. 

In the proof, we directly show the equivalence between statements (ii) and (iii) using the fact that TTC selects the core allocation for every preference profile.
To establish the equivalence between statement (i) and the others, we show that any mechanism $\varphi: \mathcal{P}^N \to \mathcal{A}$ is onto if it is individually rational and pair efficient (Lemma~\ref{lem:onto}). 

We then consider the single-dipped domain. Since TTC is not onto on this domain, the equivalence between statements (ii) and (iii) does not survive.
However, the next theorem shows that the characterization via pair efficiency is still valid on the smaller domain. 

\begin{theorem}
\label{thm:TTC_dipped}
    Let $\varphi : \check{\mathcal{P}}^N \to \mathcal{A}$ be a mechanism.
    Then $\varphi$ is individually rational, pair efficient, and NOGM if and only if it is TTC. 
\end{theorem}

\citet{hu2024characterization} showed that on the single-dipped domain, TTC is characterized by individual rationality, pair efficiency, and strategy-proofness (see also \citet{tamura2023object}, which characterizes TTC using Pareto efficiency). 
Our theorem shows that their characterization remains valid when strategy-proofness is replaced with NOGM.

\section{Conclusion}
\label{sec:con}

This paper has studied obvious manipulations in Shapley--Scarf housing markets. 
We first showed that NOM, introduced by \citet{troyan2020obvious}, is too weak in this environment: 
under a mild domain assumption, every individually rational and pair efficient mechanism is NOM, and hence NOM cannot distinguish among reasonable mechanisms. 
To address this weakness, we introduced a group version of NOM, called NOGM.
We showed that, on the full domain, TTC is the unique mechanism that is individually rational, pair efficient, and NOGM, and that pair efficiency can be replaced with ontoness. 
We also showed that the characterization via pair efficiency remains valid on the single-dipped domain, where TTC is not onto. 
These results imply that strategy-proofness in the characterizations of \citet{ekici2024pair} and \citet{hu2024characterization}, as well as group strategy-proofness in that of \citet{takamiya2001coalition}, can be replaced with NOGM. 
Taken together, they suggest that once coalitional deviations are taken into account, the NOM-like condition is strong enough to single out TTC.

Appendix~\hyperref[app:A]{A} shows that, under a richness condition, NOGM is a natural extension of NOM in a general framework and can be characterized as group strategy-proofness under mechanism uncertainty.
We leave for future work the study of the implications of NOGM in other frameworks, such as school choice and auctions.

\section*{APPENDIX A: General Characterization}
\renewcommand{\thesubsection}{A.\arabic{subsection}}
\setcounter{subsection}{0}
\label{app:A}

In this part, we provide a characterization of NOGM in a general framework. 
This result highlights that our formalization is a natural extension of NOM in \citet{troyan2020obvious}.
Note that this part uses notation different from that in the main part.

Let $N$ be the finite set of \textit{agents}.
Let $X$ be the finite set of \textit{outcomes}.
Each agent $i\in N$ has a \textit{type} $\theta_i$ and let $\Theta_i$ be the set of types of $i$.  
The value $u_i(x; \theta_i)$ denotes agent $i$’s \textit{utility} for
outcome $x$ when her type is $\theta_i$. 
Let $\Theta_N=\times_{i\in N}\Theta_i$ be the \textit{type space}.
For each $S\subseteq N$, let $\Theta_S=\times_{i\in S}\Theta_i$ and $\Theta_{-S}=\Theta_{N\setminus S}$; the notation $\theta_S$, $\theta_{-S}$, and $(\theta'_S,\theta_{-S})$ is defined analogously to the main text.
A \textit{mechanism} is a function $\varphi:\Theta_N\to X$. 

\begin{definition}
    Given a mechanism $\varphi$, $S\subseteq N$, and $\theta_S \in \Theta_S$, we say that $\theta'_S\in \Theta_S$ is a \textbf{group manipulation} for $(\theta_S,\varphi )$ if there exists $\theta_{-S}\in\Theta_{-S}$ such that
    $u_i(\varphi(\theta'_S,\theta_{-S});\theta_i)
    \ge u_i(\varphi(\theta_S,\theta_{-S});\theta_i)$ for all $i\in S$ 
    and $u_j(\varphi(\theta'_S,\theta_{-S});\theta_j) > u_j(\varphi(\theta_S,\theta_{-S});\theta_j)$ for some $j\in S$.
\end{definition}

Note that the group strategy-proofness of $\varphi$ can be defined as the condition that for each $S\subseteq N$ and each $\theta_S, \theta'_S\in \Theta_S$, $\theta'_S$ is not a group manipulation for $(\theta_S, \varphi)$. 

Here, we introduce NOGM in this general setup. 
Given a mechanism $\varphi$, $S\subseteq N$, and $\theta_S, \theta'_S\in \Theta_S$, let $B^\varphi_{\theta_S} (\theta'_S)$ be the set of \textbf{best outcomes} $b\in \{\varphi (\theta'_S, \theta_{-S} ) \mid \theta_{-S} \in \Theta_{-S} \}$ such that 
there is no $\theta'_{-S}\in \Theta_{-S}$ with $u_i (\varphi(\theta'_S, \theta'_{-S}); \theta_i) \geq u_i (b; \theta_i)$ for all $i\in S$ and $u_j (\varphi(\theta'_S, \theta'_{-S}); \theta_j) > u_j (b; \theta_j)$ for some $j\in S$.
Similarly, let $W^\varphi_{\theta_S} (\theta'_S)$ be the set of \textbf{worst outcomes} $w\in \{\varphi (\theta_S', \theta_{-S} ) \mid \theta_{-S} \in \Theta_{- S} \}$ such that 
there is no $\theta'_{-S}\in \Theta_{-S}$ with $u_i (w; \theta_i) \geq u_i (\varphi(\theta'_S, \theta'_{-S}); \theta_i)$ for all $i\in S$ and $u_j (w; \theta_j) > u_j (\varphi(\theta'_S, \theta'_{-S}); \theta_j)$ for some $j\in S$.

\begin{definition}
    Given a mechanism $\varphi$, $S\subseteq N$, and  $\theta_S\in \Theta_S$,  we say that $\theta'_S\in \Theta_S$ is an \textbf{obvious group manipulation} for $(\theta_S,\varphi)$ if either 
    \begin{itemize}
        \item[(i)] there exists $b'\in B^\varphi_{\theta_S}(\theta'_S)$ such that, for all $b\in B^\varphi_{\theta_S}(\theta_S)$, 
    $u_i(b';\theta_i)\ge u_i(b;\theta_i)$ for all $i\in S$ and $u_j(b';\theta_j)> u_j(b;\theta_j)$ for some $j\in S$; or
    \item[(ii)] there exists $w\in W^\varphi_{\theta_S}(\theta_S)$ such that, for all $w'\in W^\varphi_{\theta_S}(\theta'_S)$,  
    $u_i(w';\theta_i)\ge u_i(w;\theta_i)$ for all $i\in S$ and 
    $u_j(w';\theta_j ) > u_j(w;\theta_j)$ for some $j\in S$.
    \end{itemize}
   
\end{definition}

For $S\subseteq N$, let $\Pi_S$ be the coarsest partition of $X$ such that, for all $\pi\in\Pi_S$, all $x,y\in\pi$, and all $\theta_S\in\Theta_S$, $ u_i(x;\theta_i)=u_i(y;\theta_i)$ for all $i\in S$.
Given $\varphi$ and $\theta_S$, let $\Pi^\varphi_S(\theta_S)$ denote the set of elements $\pi\in \Pi_S$ that are attainable when coalition $S$ reports $\theta_S$, that is,
\begin{equation}
    \Pi^\varphi_S(\theta_S)
=
\bigl\{
\pi\in\Pi_S \mid
\varphi(\theta_S,\theta_{-S})\in\pi
\text{ for some }\theta_{-S}\in\Theta_{-S}
\bigr\}.
\end{equation}
Mechanisms $\varphi$ and $\psi$ are \textbf{$\bm S$-indistinguishable} if $\Pi^\varphi_{S}(\theta_S) = \Pi^\psi_{S}(\theta_S)$ for all $\theta_S\in \Theta_S$.
The type space $\Theta_N$ is \textit{rich} if for each $S\subsetneq N$,  $|\Pi_S|\le|\Theta_{-S}|$.

The following theorem characterizes obvious group manipulations.

\begin{theorem}
\label{thm:chara}
Suppose $\Theta_N$ is rich.
For each mechanism $\varphi$, each $S\subseteq N$, and each  $\theta_S,\theta'_S\in \Theta_{S}$,
$\theta'_S$ is an obvious group manipulation  for $(\theta_S,\varphi)$ if and only if for every mechanism $\psi$ that is $S$-indistinguishable from $\varphi$, $\theta'_S$ is a group manipulation for $(\theta_S,\psi)$.
\end{theorem}

\begin{proof}
Let $\varphi$ be a mechanism, let $S\subseteq N$, and let $\theta_S,\theta'_S\in \Theta_{S}$.
If $\theta_S = \theta'_S$, then $\theta'_S$ is neither an obvious group manipulation for $(\theta_S, \varphi)$ nor a group manipulation for $(\theta_S, \varphi)$; since $\varphi$ is $S$-indistinguishable from itself, the equivalence holds trivially.
Hence, we assume $\theta_S \neq \theta'_S$.

First, we consider the case with $S = N$. 
Take any mechanism $\psi$ that is $N$-indistinguishable from $\varphi$.
If $\theta'_N$ is an obvious group manipulation for $(\theta_N,\varphi)$, then $B^\varphi_{\theta_N}(\theta_N)=W^\varphi_{\theta_N}(\theta_N)=\{\varphi(\theta_N)\}$,  $B^\varphi_{\theta_N}(\theta'_N)=W^\varphi_{\theta_N}(\theta'_N)=\{\varphi(\theta'_N)\}$, $u_i(\varphi(\theta_N);\theta_i)=u_i(\psi(\theta_N);\theta_i)$ for all $i\in N$, and $u_i(\varphi(\theta'_N);\theta_i)=u_i(\psi(\theta'_N);\theta_i)$ for all $i\in N$ imply that $u_i( \psi(\theta'_N) ;\theta_i)\ge u_i(\psi(\theta_N);\theta_i)$ for all $i\in N$ and $u_j(\psi(\theta'_N);\theta_j)> u_j(\psi(\theta_N);\theta_j)$ for some $j\in N$.
Thus, $\theta'_N$ is a group manipulation for $(\theta_N,\psi)$.
Conversely, suppose that $\theta'_N$ is a group manipulation for $(\theta_N, \psi)$ for every mechanism $\psi$ that is $N$-indistinguishable from $\varphi$.
In particular, $\theta'_N$ is a group manipulation for $(\theta_N, \varphi)$.
Since $B^\varphi_{\theta_N}(\theta_N) = W^\varphi_{\theta_N}(\theta_N) = \{\varphi(\theta_N)\}$ and $B^\varphi_{\theta_N}(\theta'_N) = W^\varphi_{\theta_N}(\theta'_N) = \{\varphi(\theta'_N)\}$, $\theta'_N$ is an obvious group manipulation for $(\theta_N, \varphi)$.

Next, we consider the case with $S\subsetneq N$.
Assume that $\theta'_S$ is an obvious group manipulation for $(\theta_S,\varphi)$. 
Then, at least one of the following conditions holds:
\begin{itemize}
\item [(i)] there exists $b'\in B^\varphi_{\theta_S}(\theta'_S)$ such that, for all  $b\in B^\varphi_{\theta_S}(\theta_S)$,
$u_i(b';\theta_i)\ge u_i(b;\theta_i)$ for all $i\in S$ and $u_j(b';\theta_j)> u_j(b;\theta_j)$ for some $j\in S$;
or
\item [(ii)] there exists $w\in W^\varphi_{\theta_S}(\theta_S)$ such that, for all $w'\in W^\varphi_{\theta_S}(\theta'_S)$,  
$u_i(w';\theta_i)\ge u_i(w;\theta_i)$ for all $i\in S$ and 
$u_j(w';\theta_j) > u_j(w;\theta_j)$ for some $j\in S$.
\end{itemize}
Take any mechanism $\psi$ that is $S$-indistinguishable from $\varphi$.

If condition (i) holds, then by $\Pi^\varphi_S(\theta'_S)=\Pi^\psi_S(\theta'_S)$, there exists $\theta_{-S}$ such that, for all $i \in S$, $u_i(\psi(\theta'_S,\theta_{-S});\theta_i)=u_i(b';\theta_i)$.
By $\Pi^\varphi_S(\theta_S)=\Pi^\psi_S(\theta_S)$, the finiteness of $X$, and the definition of $B^\varphi_{\theta_S}(\theta_S)$, there exists $b^*\in B^\varphi_{\theta_S}(\theta_S)$ such that, for all $i \in S$, $u_i(b^*;\theta_i)\ge u_i(\psi(\theta_S,\theta_{-S});\theta_i)$. 
Then, it follows from condition (i) that for all $i\in S$, 
\begin{equation}
    u_i(\psi(\theta'_S,\theta_{-S});\theta_i)
    =u_i(b';\theta_i)
    \ge u_i(b^*;\theta_i)
    \ge u_i(\psi(\theta_S,\theta_{-S});\theta_i)
\end{equation}
and that for some $j\in S$,
\begin{equation}
     u_j(\psi(\theta_S',\theta_{-S});\theta_j)
    =u_j(b';\theta_j)
    > u_j(b^*;\theta_j)
    \ge
    u_j(\psi(\theta_S,\theta_{-S});\theta_j).
\end{equation}
Thus, $\theta'_S$ is a group manipulation for $(\theta_S,\psi)$.

On the other hand, if condition (ii) holds, then by $\Pi^\varphi_S(\theta_S)=\Pi^\psi_S(\theta_S)$, there exists $\theta_{-S}$ such that, for all $i \in S$, $u_i(\psi(\theta_S,\theta_{-S});\theta_i)=u_i(w;\theta_i)$. 
By $\Pi^\varphi_S(\theta'_S)=\Pi^\psi_S(\theta'_S)$, the finiteness of $X$, and the definition of $W^\varphi_{\theta_S}(\theta'_S)$, there exists $w^* \in W^\varphi_{\theta_S}(\theta'_S)$ such that, for all $i \in S$, $u_i(\psi(\theta'_S,\theta_{-S});\theta_i)\ge u_i(w^*;\theta_i)$. 
Then, it follows from condition (ii) that for all $i\in S$, 
\begin{equation}
    u_i(\psi(\theta'_S,\theta_{-S});\theta_i)
    \ge u_i(w^*;\theta_i)
    \ge u_i(w;\theta_i)
    = u_i(\psi(\theta_S,\theta_{-S});\theta_i) 
\end{equation}
and that for some $j\in S$,
\begin{equation}
    u_j(\psi(\theta'_S,\theta_{-S});\theta_j)
    \ge u_j(w^*;\theta_j)
    > u_j(w;\theta_j)
    = u_j(\psi(\theta_S,\theta_{-S});\theta_j).
\end{equation}
Thus, $\theta'_S$ is a group manipulation for $(\theta_S,\psi)$.

To show the converse, we prove the contrapositive. Assume that $\theta'_S$ is not an obvious group manipulation for $(\theta_S,\varphi)$.
We construct a mechanism $\psi$ that is $S$-indistinguishable from $\varphi$ as follows: 
First, for each $\theta^*_S\notin \{\theta_S,\theta'_S\}$ and each $\theta_{-S}\in\Theta_{-S}$,
let $\psi(\theta_S^*,\theta_{-S})=\varphi(\theta_S^*,\theta_{-S})$.
For each $\pi\in \Pi_S^\varphi(\theta_S)\cup \Pi_S^\varphi(\theta'_S)$, take $\theta^{\pi}_{-S} \in \Theta_{- S}$ that are distinct from each other.
Note that since $|\Pi_S^\varphi(\theta_S)\cup\Pi_S^\varphi(\theta'_S)|\le |\Pi_S|$ and $|\Pi_S|\le|\Theta_{-S}|$ by supposition, we have $|\Pi_S^\varphi(\theta_S)\cup\Pi_S^\varphi(\theta'_S)|\le |\Theta_{-S}|$ so we can take such $\{ \theta^{\pi}_{-S} \}_{{\pi} \in \Pi_S^\varphi(\theta_S)\cup\Pi_S^\varphi(\theta'_S)}$. 
For each $\pi\in \Pi_S^\varphi(\theta_S)\cup \Pi_S^\varphi(\theta'_S)$,  choose and fix an outcome $x\in\pi$.
\begin{itemize}
    \item Suppose $\pi\in \Pi_S^\varphi(\theta_S)\setminus\Pi_S^\varphi(\theta'_S)$.
Since $\pi\in\Pi_S^\varphi(\theta_S)$ and $X$ is finite, there exists $w_\pi\in W^\varphi_{\theta_S}(\theta_S)$ such that
$u_i(x;\theta_i)\ge u_i(w_\pi;\theta_i)$ for all $i\in S$.
By assumption, there exists $w'_\pi\in W^\varphi_{\theta_S}(\theta'_S)$ such that $u_i(w_\pi;\theta_i)>u_i(w'_\pi;\theta_i)$ for some $i\in S$, or $u_j(w_\pi;\theta_j)\ge u_j(w'_\pi;\theta_j)$ for all $j\in S$.
Consequently, $u_i(x;\theta_i)>u_i(w'_\pi;\theta_i)$ for some $i\in S$, or $u_j(x;\theta_j)\ge u_j(w'_\pi;\theta_j)$ for all $j\in S$.
\item Suppose $\pi \in \Pi_S^\varphi(\theta'_S)\setminus\Pi_S^\varphi(\theta_S)$.
Since $\pi\in\Pi_S^\varphi(\theta'_S)$ and $X$ is finite, there exists $b'_\pi\in B^\varphi_{\theta_S}(\theta'_S)$ such that $u_i(b'_\pi;\theta_i)\ge u_i(x;\theta_i)$ for all $i\in S$.
By assumption, there exists $b_{\pi}\in B^\varphi_{\theta_S}(\theta_S)$ such that $u_i(b_{\pi};\theta_i)>u_i(b'_\pi;\theta_i)$ for some $i\in S$, or $u_j(b_{\pi};\theta_j)\ge u_j(b'_\pi;\theta_j)$ for all $j\in S$.
Consequently, $u_i(b_{\pi};\theta_i)>u_i(x;\theta_i)$ for some $i\in S$, or $u_j(b_{\pi};\theta_j)\ge u_j(x;\theta_j)$ for all $j\in S$.
\end{itemize}

For each $\pi\in \Pi_S^\varphi(\theta_S)\setminus\Pi_S^\varphi(\theta'_S)$, let $\psi(\theta_S,\theta^\pi_{-S})=x$ and $\psi(\theta'_S,\theta^{\pi}_{-S})=w'_{\pi}$; for each $\pi\in \Pi_S^\varphi(\theta_S')\setminus\Pi_S^\varphi(\theta_S)$, let $\psi(\theta_S,\theta^{\pi}_{-S})=b_{\pi}$ and $\psi(\theta'_S,\theta^{\pi}_{-S})=x$; and, for each $\pi\in \Pi_S^\varphi(\theta'_S)\cap\Pi_S^\varphi(\theta_S)$, let $\psi(\theta_S,\theta^{\pi}_{-S})=\psi(\theta'_S,\theta^{\pi}_{-S})=x$.
Take any $\hat{b}\in B^\varphi_{\theta_S}(\theta_S)$ and $\hat{b}'\in B^\varphi_{\theta_S}(\theta'_S)$ such that $u_i(\hat{b};\theta_i)>u_i(\hat{b}';\theta_i)$ for some $i\in S$ or $u_j(\hat{b};\theta_j)\ge u_j(\hat{b}';\theta_j)$ for all $j\in S$ (note that we can take such outcomes since $\theta'_S$ is not an obvious group manipulation for $(\theta_S, \varphi)$).
Finally, for each remaining $\theta_{-S} \in \Theta_{-S}$, let $\psi(\theta_S,\theta_{-S})=\hat{b}$ and  $\psi(\theta'_S,\theta_{-S})=\hat{b}'$. 
Note that $S$-indistinguishability follows from  $\Pi_S^\psi(\theta_S)=\Pi_S^\varphi(\theta_S)$, $\Pi_S^\psi(\theta'_S)=\Pi_S^\varphi(\theta'_S)$ and $\Pi_S^\psi(\theta^*_S)=\Pi_S^\varphi(\theta^*_S)$ for all $\theta^*_S\notin \{\theta_S,\theta'_S\}$.
Here, by the construction of $\psi$, for each $\theta_{-S} \in \Theta_{-S}$, we have 
$u_i(\psi(\theta_S,\theta_{-S});\theta_i)> u_i(\psi(\theta'_S,\theta_{-S});\theta_i)$ for some $i\in S$ 
or $u_j(\psi(\theta_S,\theta_{-S});\theta_j)\ge u_j (\psi(\theta'_S,\theta_{-S});\theta_j)$ for all $j\in S$.
Thus, $\theta'_S$ is not a group manipulation for $(\theta_S,\psi)$.
\end{proof}

The richness condition is mild.
For example, the housing market model studied in the main part satisfies this condition on the full domain.
Indeed, for any $S\subsetneq N$, we have $|\Pi_S|=|\{a_S\mid a\in\mathcal A\}|=\frac{n!}{(n-|S|)!}\le n!\le(n!)^{n-|S|}=|\mathcal P^{N\setminus S}|$.

Theorem~\ref{thm:chara} extends the characterization of obvious manipulations in  \cite{troyan2020obvious} to obvious group manipulations.
It also generalizes their result in two further respects.
First, we dispense with the assumption that $n\ge3$.
Thus, this assumption can be removed from Theorem~1 of \cite{troyan2020obvious}.
Second, we work in a more general model in which outcomes need not be represented as profiles of individual allocations.

Theorem \ref{thm:chara} shows that a deviation is an obvious group manipulation for a mechanism if and only if it is a group manipulation under every mechanism indistinguishable from the coalition.
Following the interpretation in \cite{troyan2020obvious}, obvious group manipulations can also be understood as manipulations that a group can identify with only partial information about the mechanism, namely, the sets of outcomes attainable from each group report.
We say that a mechanism $\varphi$ is \textbf{group strategy-proof under mechanism uncertainty} if, for all $S\subseteq N$ and all $\theta_S\in\Theta_S$, there is no $\theta'_S\in\Theta_S$ that is a group manipulation for $(\theta_S, \psi)$ for every mechanism $\psi$ that is $S$-indistinguishable from $\varphi$.

Theorem~\ref{thm:chara} immediately yields the following equivalence.

\begin{corollary}
Suppose $\Theta_N$ is rich.
A mechanism is NOGM if and only if it is group strategy-proof under mechanism uncertainty.
\end{corollary}

\section*{APPENDIX B: Omitted Proofs}
\renewcommand{\thesubsection}{B.\arabic{subsection}}
\setcounter{subsection}{0}
\label{app:B}

\subsection{Proof of Proposition~\ref{prop:non-restriction}}

Suppose that $\mathcal{D}$ satisfies top-attainability or bottom-attainability.
Let $\varphi: \mathcal{D}^N\to \mathcal{A}$ be an  individually rational and pair efficient mechanism. 

First, we consider the case where $\mathcal{D}$ satisfies top-attainability. 
Take any agent $i\in N$.
Fix any $\succ_i\in \mathcal{D}$.
We show that agent $i$ cannot manipulate the worst case.
Since $\varphi$ is individually rational, $w_i \succeq_i o_i$ for all $w\in W^{\varphi}_{\succ_i}(\succ_i)$.
Thus it is sufficient to show that, for any $\succ_i'\in\mathcal{D}$, we have $o_i\succeq_i w_i$ for all $w\in W_{\succ_i}^{\varphi}(\succ'_i)$.
Take any $\succ'_{i}\in\mathcal{D}$. Consider $\succ'_{-i}\in\mathcal{D}^{N\setminus \{i\}}$ such that, for each $j\in N\setminus\{i\}$, $o_j\succeq_j' o$ for all $o\in O$.
Then, by individual rationality, for each $j\in N\setminus\{i\}$, $\varphi_j(\succ')=o_j$, and hence $\varphi_i(\succ')=o_i$. 

We then show that agent $i$ cannot manipulate the best case.
Let $o^*\in O$ be such that $o^*\succeq_i o$ for all $o\in O$. 
It is sufficient to show that there exists $\succ_{-i}\in\mathcal{D}^{N\setminus \{i\}}$ such that $\varphi_i(\succ)=o^*$. 
Let $j\in N$ be such that $o_j=o^*$. Consider $\succ_{-i}\in\mathcal{D}^{N\setminus \{i\}}$ such that, for each $k\in N\setminus\{i,j\}$, $o_k\succeq_k o$ for all $o\in O$ and if $j\neq i$, $o_i\succeq_j o$ for all $o\in O$.
If $j = i$, then individual rationality implies that for each $k\in N\setminus\{i\}$, $\varphi_k(\succ)=o_k$. 
Then we have $\varphi_i(\succ)=o_i=o^*$.
On the other hand, if $j \neq i$, then individual rationality implies that for each $k\in N\setminus\{i,j\}$, $\varphi_k(\succ)=o_k$.
If $\varphi_i(\succ)=o_i$ and $\varphi_j(\succ)=o^*$, then $\varphi_j(\succ) = o^*\succ_i \varphi_i(\succ)$ and $\varphi_i (\succ) = o_i \succ_j \varphi_j(\succ)$, which contradicts pair efficiency.
Thus, we have $\varphi_i(\succ)=o^*$.

Second, we consider the case  where $\mathcal{D}$ satisfies bottom-attainability.
Take any agent $i\in N$.
Fix any $\succ_i\in \mathcal{D}$.
We show that agent $i$ cannot manipulate the worst case.
Since $\varphi$ is individually rational, $w_i\succeq_i o_i$ for all $w\in W^{\varphi}_{\succ_i}(\succ_i)$.
Thus it is sufficient to show that, for any $\succ_i'\in\mathcal{D}$, we have $o_i\succeq_iw_i$ for all $w\in W_{\succ_i}^{\varphi}(\succ'_i)$.
Take any $\succ'_{i}\in\mathcal{D}$.
Consider $\succ'_{-i}\in\mathcal{D}^{N\setminus \{i\}}$ such that, for each $j\in N\setminus\{i\}$, $o\succeq_j' o_i$ for all $o\in O$. Then, for each $j\in N\setminus \{i\}$, the individual rationality of $\varphi$ and $o_j \succ'_j o_i$ imply that  $\varphi_j(\succ')\neq o_i$, and hence $\varphi_i(\succ')= o_i$.

We then show that agent $i$ cannot manipulate the best case.
Let $o^*\in O$ be such that $o^*\succeq_i o$ for all $o\in O$.
Then it is sufficient to show that there exists $\succ_{-i}\in\mathcal{D}^{N\setminus \{i\}}$ such that $\varphi_i(\succ)=o^*$. 
Let $j\in N$ be such that $o_j=o^*$.
Consider $\succ_{-i}\in \mathcal{D}^{N\setminus\{i\}}$ such that, for each $k\in N\setminus\{i\}$, it holds that $o\succeq_k o^*$ for all $o\in O$.
If $j=i$, then for each $k\in N\setminus\{i\}$, $o_k\neq o^*(=o_i)$ implies $o_k\succ_k o^*$.
By individual rationality, $\varphi_k(\succ)\succeq_k o_k\succ_k o^*$, so $\varphi_k(\succ)\neq o^*$.
Since $\varphi(\succ)$ is a bijection, we have $\varphi_i(\succ)=o^*$.
On the other hand, if $j\neq i$, then for each $k\in N\setminus\{i,j\}$, the same argument as above (using $o_k\neq o^*$) gives $\varphi_k(\succ)\neq o^*$.
Thus, either $\varphi_i(\succ)=o^*$  or $\varphi_j(\succ)=o^*$ holds. 
If $\varphi_j(\succ)=o^*$, then we have $\varphi_j(\succ) = o^* \succ_i \varphi_i(\succ)$ and $\varphi_i(\succ) \succ_j o^* = \varphi_j(\succ)$, which contradicts pair efficiency. 
Thus, we have $\varphi_i(\succ)=o^*$. 
\qed

\subsection{Proof of Theorem~\ref{thm:TTC_full}}

It is straightforward to see that TTC satisfies the properties in the statement.
In particular, TTC is NOGM because it is group strategy-proof and group strategy-proofness implies NOGM
(see Appendix~\hyperref[app:C]{C}).  
First, we prove that (i) implies (ii). 
This part follows from the next lemma. 

\begin{lemma}
\label{lem:onto}
    If a mechanism $\varphi: \mathcal{P}^N \to \mathcal{A}$ is individually rational and pair efficient, then it is onto.
\end{lemma}

\begin{proof}
    Let $\varphi: \mathcal{P}^N \to \mathcal{A}$ be a mechanism that is individually rational and pair efficient. 
    Take any allocation $a\in \mathcal{A}$. 
    It is sufficient to prove that there exists $\succ \in \mathcal{P}^N$ such that $a$ is the unique individually rational and pair efficient allocation.

    Let $\mathcal{G} = \{ G_1, \ldots, G_k \}$ be the finest partition such that $\{o_i \in O \mid i\in G_\ell\} = \{a_i \in O \mid i\in G_\ell\}$ for each $\ell \in \{1,\ldots, k\}$. 
    Consider a preference profile $\succ$ such that, for each $\ell \in \{1,\ldots, k\}$, (i) if $G_\ell = \{i\}$ for some $i\in N$, then $o_i \succeq_i o$ for all $o\in O$; 
    (ii) if $G_\ell = \{i, j\}$ for some distinct $i, j\in N$, then $o_j \succ_i o_i \succ_i o$ and $o_i \succ_j o_j \succ_j o$ for all $o\in O\setminus \{ o_i, o_j \}$; 
    and (iii) if $|G_\ell| = m$ for some $m\geq 3$, then, letting $G_\ell = \{1,2,\ldots, m\}$ and $(a_1,a_2, a_3, \ldots, a_{m-1}, a_m) = (o_2, o_3, o_4, \ldots,o_m, o_1)$ without loss of generality, the following holds: 
    \begin{table}[H]
    \centering
    \begin{tabular}{cccccccc}
    $1$ & $2$ & $3$ & $4$ & $\cdots$ & $m-2$ & $m-1$ & $m$ \\
    \hline
    $o_2$ & $o_3$ & $o_4$ & $o_5$ & $\cdots$ & $o_{m-1}$ & $o_{m}$   & $o_1$ \\
    $o_1$ & $o_2$ & $o_2$ & $o_2$ & $\cdots$ & $o_2$   & $o_2$   & $o_2$ \\
       $\vdots$  &  $\vdots$    & $o_3$ & $o_4$ & $\cdots$ & $o_{m-2}$   & $o_{m-1}$   & $o_m$ \\
        &     &  $\vdots$   &   $\vdots$  & $\cdots$ & $\vdots$ & $\vdots$ & $\vdots$ \\
    \hline
    \end{tabular}
    \caption{Preference profile $\succ$ in $G_\ell$}
    \label{tab:preference_profile}
    \end{table}
    It is clear that $a$ is individually rational and pair efficient at $\succ$. 
    Let $a' \in \mathcal{A}$ be such that it is individually rational and pair efficient at $\succ$. 
    Let $\ell \in \{1,\ldots, k\}$. 
    (i) If $G_\ell = \{i\}$ for some $i\in N$, then $a'_i = o_i = a_i$ by the individual rationality of $a'$. 
    (ii) If $G_\ell = \{i, j\}$ for some distinct $i, j\in N$, then by pair efficiency and individual rationality, we have $a'_i = o_j = a_i$ and $a'_j = o_i =  a_j$. 
    (iii) If $|G_\ell| = m$ for some $m\geq 3$, then let $G_\ell = \{1,2,\ldots, m\}$ and $(a_1,a_2, a_3, \ldots, a_{m-1}, a_m) = (o_2, o_3, o_4, \ldots,o_m, o_1)$ again,  without loss of generality. In this case, $(a'_1, a'_2)$ is one of $(o_2, o_3)$, $(o_1, o_2)$, and $(o_1, o_3)$. 
    \begin{itemize}
        \item[(iii-a)] If $(a'_1, a'_2) = (o_2, o_3)$, we show that $a'_k = o_{k+1}$ for $k = 3, \ldots, m-1$ and $a'_m = o_1$.
        For each $k \in \{3, \ldots, m-1\}$, the objects that agent $k$ weakly prefers to $o_k$ are $o_{k+1}$, $o_2$, and $o_k$.
        Since $a'_1 = o_2$ and $a'_{k-1} = o_k$ (by $a'_2 = o_3$ if $k = 3$ and by induction otherwise), individual rationality implies $a'_k = o_{k+1}$.
        Similarly, the objects that agent $m$ weakly prefers to $o_m$ are $o_1$, $o_2$, and $o_m$, and since $a'_1 = o_2$ and $a'_{m-1} = o_m$, we have $a'_m = o_1$.
        Thus, $(a'_1, a'_2, \ldots, a'_m) = (a_1, a_2, \ldots, a_m)$.

        \item[(iii-b)] If $(a'_1, a'_2) = (o_1, o_2)$, then by the individual rationality of $a'$, we have $a'_m = o_m$,  $a'_{m-1} = o_{m-1}$, $\cdots$, and $a'_3 = o_3$. 
        Since $a'_2 = o_2 \succ_3 o_3 = a'_3$ and $a'_3 = o_3 \succ_2 o_2 = a'_2$, $a'$ is not pair efficient. 

        \item[(iii-c)] If $(a'_1, a'_2) = (o_1, o_3)$, then  there exists $j\in \{3,4,\ldots, m\}$ such that $a'_{j} = o_2$. 
        Assume $3 \leq j \leq m-1$.
        Then, by $\{i \in N \mid o_{j+1} \succeq_i o_i\} = \{j , j+1 \}$ and the individual rationality of $a'$, we have $a'_{j + 1} = o_{j+1}$. 
        Since $a'_{j+1} = o_{j+1} \succ_j o_2 = a'_j$ and $a'_j = o_2 \succ_{j+1} o_{j+1} = a'_{j+1}$, it contradicts pair efficiency. 
        Thus, $a'_m = o_2$. This is also a contradiction to pair efficiency because $a'_{m} = o_{2} \succ_1 o_1 = a'_1$ and $a'_1 = o_1 \succ_m o_{2} = a'_{m}$. 
    \end{itemize}
    Since this argument can be applied to every $G\in \mathcal{G}$, we have $a' = a$. 
    Thus, $a$ is the unique allocation that is individually rational and pair efficient at $\succ$.  
    Hence $\varphi(\succ) = a$, so $\varphi$ is onto.
\end{proof}

We then prove that (ii) implies (iii). 

\begin{lemma}
\label{lem:PO}
Any onto and NOGM mechanism $\varphi: \mathcal{P}^N \to\mathcal{A}$ is Pareto efficient.
\end{lemma}

\begin{proof}
Let $\varphi$ be an onto and NOGM mechanism. 
Suppose to the contrary that for some $\succ\in \mathcal{P}^N$, there exists $a \in \mathcal{A}$ such that $a_i \succeq_i \varphi_i(\succ)$ for all $i \in N$ and $a_j\succ_j\varphi_j(\succ)$ for some $j \in N$.
Since $\varphi$ is onto, there exists $\succ'$ with $\varphi(\succ') = a$. 
Then, we have $\varphi_i(\succ') \succeq_i \varphi_i(\succ)$ for all $i \in N$ and $\varphi_j(\succ') \succ_j\varphi_j(\succ)$ for some $j \in N$.
Since $W^\varphi_{\succ}(\succ)=\{\varphi(\succ)\}$ and $W^\varphi_{\succ}(\succ')=\{\varphi(\succ')\}$, it contradicts the assumption that $\varphi$ is NOGM.
Hence, $\varphi$ is Pareto efficient.
\end{proof}

Let $\varphi$ be a mechanism that is individually rational, onto, and NOGM.
Suppose to the contrary that $\varphi$ is not TTC.
Then, since the TTC allocation is the unique core allocation in each preference profile (cf.~Section~\ref{sec:model}), there exist $S\subseteq N$, $\succ\in \mathcal{P}^N$ and $a\in \mathcal{A}$ such that 
$\{o_i\mid i\in S\}=\{a_i\mid i \in S\}$, $a_i\succeq _i\varphi_i(\succ)$ for all $i\in S$, and $a_j\succ _j\varphi_j(\succ)$ for some $j\in S$.
Let $w\in W^{\varphi}_{\succ_S}(\succ_S)$ be such that for each $i\in S$, $\varphi_i(\succ)\succeq_i w_i$.
Then, $a_i\succeq _iw_i$ for all $i\in S$ and $a_j\succ_jw_j$ for some $j\in S$.
Consider $\succ'_S\in \mathcal{P}^S$ such that for each $i\in S$ and each $\hat o\in O\setminus\{a_i,o_i\}$, $a_i\succeq'_io_i\succ'_i\hat o$.

Then, for each $\hat\succ_{-S}\in\mathcal{P}^{N\setminus S}$, $\{\varphi_i(\succ'_S,\hat \succ_{-S} )\mid i\in S\}=\{a_i\mid i\in S\}$
since $\varphi$ is individually rational,
and it follows that $\varphi _i(\succ'_S,\hat \succ_{-S} )=a_i$ for each $i\in S$ since $\varphi$ is Pareto efficient (Lemma~\ref{lem:PO}).
Hence, for each $w'\in W^\varphi_{\succ_S}(\succ'_S)$, $w'_i=a_i$ for each $i\in S$.
Since $a_i\succeq _iw_i$ for all $i\in S$ and $a_j\succ _jw_j$ for some $j\in S$, it holds that for each $w'\in W^\varphi_{\succ_S}(\succ'_S)$,  
$w'_i\succeq_iw_i$ for all $i\in S$ and $w'_j\succ_jw_j$ for some $j\in S$, contradicting that $\varphi$ is NOGM. 
Therefore, $\varphi$ is TTC.
\qed

\subsection{Proof of Theorem~\ref{thm:TTC_dipped}}

We omit the proof for the if part. Let $\varphi : \check{\mathcal{P}}^N \to \mathcal{A}$ be a mechanism that is pair efficient, individually rational, and NOGM. 
Let $\succ \in \check{\mathcal{P}}^N$. 
    
We first consider agents $1$ and $n$. 
By the definition of single-dippedness, both agents prefer $o_1$ or $o_n$ most. 
If each of them prefers their endowments most,
then by individual rationality, $\varphi_1 (\succ) = o_1$ and $\varphi_n (\succ) = o_n$. Let $N^1 = \{1,n\}$ and $O^1 = \{ o_1, o_n\}$. 
If only one agent $i \in \{1, n\}$ prefers their endowment $o_i$ most,
then  by individual rationality, $\varphi_i (\succ) = o_i$. 
In this case, let $N^1 = \{i\}$ and  $O^1 = \{ o_i \}$.
If agent $1$ prefers $o_n$ most and agent $n$ prefers $o_1$ most,
then consider $\succ'_1, \succ'_n \in \check{\mathcal{P}}$ such that for all $o\in O\setminus \{o_1, o_n\}$,
$o_n \succ'_1 o_1 \succ'_1 o$ and $o_1 \succ'_n o_n \succ'_n o$. 
By pair efficiency and individual rationality,
for all $\succ'_{N\setminus \{1,n\}}$,
we have $\varphi_1 (\succ') = o_n$ and $\varphi_n (\succ') = o_1$. 
Since $\varphi$ is NOGM, $\varphi_1 (\succ) = o_n$ and $\varphi_n (\succ) = o_1$.
In this case, let $N^1 = \{1,n\}$ and $O^1 = \{ o_1, o_n\}$. 
Note that the above argument does not depend on the preferences of the agents in $N \setminus N^1$. Hence, at every profile in which the agents in $N^1$ report $\succ_{N^1}$, they receive the objects in $O^1$ as specified above.

If $N \setminus N^1$ consists of a single agent $i$, then $O \setminus O^1 = \{o_i\}$, and since the objects in $O^1$ are assigned to the agents in $N^1$, we have $\varphi_i(\succ) = o_i$, which coincides with the object that TTC assigns to $i$.
Otherwise, consider two distinct agents $i^*, i_* \in N \setminus N^1$ such that $i_* \le i \le i^*$ for all $i \in N \setminus N^1$.
By the definition of single-dippedness, both agents prefer $o_{i^*}$ or $o_{i_*}$ most within the remaining objects.
If each of them prefers their endowments most within the remaining objects,
then individual rationality implies
$\varphi_{i^*} (\succ) = o_{i^*}$ and $\varphi_{i_*} (\succ) = o_{i_*}$. Let $N^2 = \{i^*,i_*\}$ and $O^2 = \{o_{i^*},o_{i_*}\}$. 
If only one agent $i \in \{i^*,i_*\}$ prefers their endowment $o_i$ most within the remaining objects,
then  by individual rationality, $\varphi_i (\succ) = o_i$. In this case, let $N^2 = \{i\}$ and $O^2 = \{o_i\}$.
If agent $i^*$ prefers $o_{i_*}$ most and agent $i_*$ prefers $o_{i^*}$ most within the remaining objects, then 
consider $\succ'_{i_*}, \succ'_{i^*} \in \check{\mathcal{P}}$ such that
for all $o\in O\setminus (\{o_{i_*}, o_{i^*}\} \cup O^1)$,
$o_{i^*} \succ'_{i_*} o_{i_*} \succ'_{i_*} o$ and $o_{i_*} \succ'_{i^*} o_{i^*} \succ'_{i^*} o$. 
Let $S = N^1 \cup \{i_*, i^*\}$ and $\succ'_{N^1} = \succ_{N^1}$.
By the note above, pair efficiency, and individual rationality, for all $\succ'_{N\setminus S}$, the agents in $N^1$ receive the objects in $O^1$ as specified above, $\varphi_{i_*}(\succ') = o_{i^*}$, and $\varphi_{i^*}(\succ') = o_{i_*}$.
Since $\varphi$ is NOGM (applied to $S$ and $\succ'_S$), $\varphi_{i_*}(\succ) = o_{i^*}$ and $\varphi_{i^*}(\succ) = o_{i_*}$.
In this case, let $N^2 = \{i_*, i^*\}$ and $O^2 = \{o_{i^*},o_{i_*}\}$.
   
By repeating this argument, we can construct $\varphi(\succ)$ (note that at each step, if only one agent remains, she receives her own endowment). It is easy to see that the allocation coincides with the one induced by TTC. \qed

\section*{APPENDIX C: Logical Relations among Properties}
\renewcommand{\thesubsection}{B.\arabic{subsection}}
\setcounter{subsection}{0}
\label{app:C}
In this appendix, we first show that strategy-proofness and NOGM are logically independent using two examples. 
In what follows, we consider the three-agent case, but it is straightforward to extend these arguments to the case with any number of agents by adding dummy agents
who always receive their endowments regardless of the reported preference profile.

\begin{example}[A strategy-proof mechanism that violates NOGM]
    Consider two allocations $a = (o_1, o_2, o_3)$ and $b = (o_1, o_3, o_2)$.
    Define a mechanism $\varphi$ by $\varphi(\succ) = a$ if $o_2 \succ_1 o_3$, and $\varphi(\succ) = b$ if $o_3 \succ_1 o_2$.
    Since agent 1 always receives $o_1$ and agents 2 and 3 have no effect on the outcome, $\varphi$ is strategy-proof.
    Then we show that $\varphi$ is not NOGM. Consider a coalition $S = \{1, 2\}$ and a true preference subprofile $\succ_S$ where $o_2 \succ_1 o_3$ and $o_3 \succ_2 o_2$.
    Under truth-telling, the chosen allocation is always $a$, so $a$ is the unique worst allocation.
    If agent 1 reports $o_3 \succ'_1 o_2$, the chosen allocation is always $b$, so $b$ is the unique worst allocation.
    Since $b_1 = a_1 = o_1$ and $b_2 = o_3 \succ_2 o_2 = a_2$, $\varphi$ is not NOGM.
\end{example}

\begin{example}[A NOGM mechanism that violates strategy-proofness]
    Consider two allocations $a=(o_1,o_2,o_3)$ and $b=(o_2,o_3,o_1)$.
    For any preference profile $\succ$, let $k(\succ) = |\{i \in N \mid b_i \succ_i a_i\}|$.
    Define a mechanism $\varphi$ as follows:
    \begin{itemize}
        \item $\varphi(\succ) = b$ if $k(\succ)$ is odd.
        \item $\varphi(\succ) = a$ if $k(\succ)$ is even.
    \end{itemize}
    
    First, we show that $\varphi$ violates strategy-proofness. 
    Consider a true preference profile $\succ$ where $o_3\succ_i o_2 \succ_i o_1$ for all $i\in N$. Since $k(\succ)=2$, we have $\varphi(\succ) = a$.
    If agent 1 reports $o_1 \succ'_1 o_2 \succ'_1 o_3$, then $k(\succ'_1,\succ_2,\succ_3)=1$, and hence $\varphi(\succ'_1,\succ_2,\succ_3) = b$. Since $\varphi_1(\succ'_1,\succ_2,\succ_3) = o_2 \succ_1 o_1 =\varphi_1(\succ)$, $\varphi$ violates strategy-proofness.
    
    Next, we show that $\varphi$ is NOGM.
    Let $S \subseteq N$ be any coalition. We consider two cases:
    
    \noindent \textbf{Case 1:} $S\subsetneq N$.
    For any report $\succ'_S$, the set of possible allocations $\{\varphi(\succ'_S, \succ_{-S}) \mid \succ_{-S} \in \mathcal{P}^{N \setminus S}\}$ is always $\{a, b\}$.
    
    \noindent \textbf{Case 2:} $S=N$.
    Take any $\succ$. If $k(\succ)=3$, $b_i \succ_i a_i$ for all $i\in N$ and $\varphi(\succ) = b$. If $k(\succ)=0$, $a_i \succ_i b_i$ for all $i\in N$ and $\varphi(\succ) = a$. Otherwise, there exist $i\in N$ with $b_i \succ_i a_i$ and $j\in N$ with $a_j \succ_j b_j$. 
    In all cases, for any $\succ'$, $\varphi(\succ')$ does not Pareto dominate $\varphi(\succ)$ at $\succ$.
    
    Therefore, $\varphi$ satisfies NOGM.
\end{example}

We next show that NOGM is weaker than group strategy-proofness.
Let $\varphi$ be a group strategy-proof mechanism on $\mathcal{D}^N$.
Take any $S\subseteq N$ and any  $\succ_S,\succ'_S\in \mathcal{D}^{S}$.
We first prove the claim for best allocations.
Take any  $b'\in B^\varphi_{\succ_S}(\succ'_S)$ and let  $\succ_{-S}\in \mathcal{D}^{N\setminus S}$ be such that $\varphi(\succ'_S,\succ_{-S})=b'$.
Since $\varphi$ is group strategy-proof, $\varphi_i(\succ_S,\succ_{-S})\succeq _i b'_i$ for all $ i\in S $ or $
\varphi_j(\succ_S,\succ_{-S})\succ _j b'_j$ for some $j\in S$.
By the finiteness of $\mathcal A$, there exists $b\in B^\varphi_{\succ_S}(\succ_S)$ such that $b_i \succeq_i\varphi_i(\succ_S,\succ_{-S})$ for all $i\in S$.
Hence, $b_i \succeq _ib'_i $ for all $ i\in S $ or $
b_j \succ _j b'_j$ for some $j\in S$.
We next prove the claim for worst allocations.
Take any $w\in W^\varphi_{\succ_S}(\succ_S)$ and let $\succ_{-S}\in \mathcal{D}^{N\setminus S}$ be such that $\varphi(\succ_S,\succ_{-S})=w$.
Since $\varphi$ is group strategy-proof, $w_i\succeq_i\varphi_i(\succ'_S,\succ_{-S}) $ for all $ i\in S $ or $ w_j\succ _j
\varphi_j(\succ'_S,\succ_{-S})$ for some $j\in S$.
By the finiteness of $\mathcal{A}$, there exists $w'\in W^\varphi_{\succ_S}(\succ'_S)$ such that $\varphi_i(\succ'_S,\succ_{-S})\succeq_i w'_i$ for all $i\in S$.
Hence, $ w_i \succeq _iw'_i $ for all $ i\in S $ or $
w_j \succ _j w'_j$ for some $j\in S$.
Therefore, $\varphi$ is NOGM.

Since NOGM and strategy-proofness are logically independent, the converse does not hold.

\bibliographystyle{econ}
\bibliography{reference}

\end{document}